\documentclass{article}
\usepackage{arxiv}

\usepackage{graphicx}
\usepackage{amssymb}
\usepackage{lineno}
\usepackage[utf8]{inputenc}
\usepackage{adjustbox,lipsum}
\usepackage{array,graphicx}
\usepackage{booktabs}
\usepackage{pifont}
\usepackage{subfigure} 
\usepackage{algorithm}
\usepackage{algpseudocode}
\usepackage{rotating}

\usepackage{url}
\usepackage{geometry}
\usepackage{float}
\usepackage{textcomp}
\usepackage{amsmath}
\usepackage{multirow}
\usepackage{booktabs}
\usepackage{pifont}
\usepackage{fontenc}
\usepackage{color}
\usepackage{subfigure} 
\usepackage[nottoc]{tocbibind}

\title{A novel auction system for selecting advertisements in Real-Time bidding}

\author{
 Luis Miralles-Pechuán \\
  School of Computing. Technological University Dublin, Ireland \\
  \texttt{luis.miralles@tudublin.ie} \\
   \And
  Fernando Jiménez \\
Department of Information and Communication Engineering,\\ University of Murcia, 30071 Murcia, Spain \\
  \texttt{fernan@um.es} \\
  \And
 José Manuel García \\
Department of Information and Communication Engineering,\\ University of Murcia, 30071 Murcia, Spain \\
  \texttt{jmgarcia@um.es} \\
}

\newcommand*\rot{\rotatebox{90}}

\begin{document}

\maketitle


\begin{abstract}
Real-Time Bidding is a new Internet advertising system that has become very popular in recent years. This system works like a global auction where advertisers bid to display their impressions in the publishers' ad slots. The most popular system to select which advertiser wins each auction is the Generalized second-price auction in which the advertiser that offers the most wins the bet and is charged with the price of the second largest bet. In this paper, we propose an alternative betting system with a new approach that not only considers the economic aspect but also other relevant factors for the functioning of the advertising system. The factors that we consider are, among others, the benefit that can be given to each advertiser, the probability of conversion from the advertisement, the probability that the visit is fraudulent, how balanced are the networks participating in RTB and if the advertisers are not paying over the market price. In addition, we propose a methodology based on genetic algorithms to optimize the selection of each advertiser. We also conducted some experiments to compare the performance of the proposed model with the famous Generalized Second-Price method. We think that this new approach, which considers more relevant aspects besides the price, offers greater benefits for RTB networks in the medium and long-term.
\end{abstract}


\section{Introduction}

Each time, more companies announce their products on the Internet since it is an effective channel to reach potential customers. This channel is very appropriate for selling intangible products such as hotel reservations, flight tickets, company shares or medical insurances anywhere in the world \cite{inter2015,lange2013war}.

There are some good reasons that explain why online advertising has seen such enormous growth. For instance, online campaigns advertisers are able to segment Internet users based on different features such as access time, city, age or gender. This means that advertisers can target their campaigns more assertively by selecting the right user profile depending on the advertised product \footnote{This technique is known as microtargeting and it significantly improves advertisers' profitability by setting up the campaigns parameter \cite{sivadas1998internet}.}.

Another advantage is that advertisers are able to know in real time their campaigns' performance, which allows them to adjust their campaigns' settings to make them more effective each time \cite{archak2010budget}.

Additionally, online campaigns are available for any budget size, from small businesses and freelancers which spend a few dollars a day to international companies that invest millions of dollars
weekly.

In the first Internet advertising model, advertisers contacted directly with publishers in order to display their banners. With the passage of time, and as more advertisers and more publishers wanted to participate, this system of managing to advertise became a bit obsolete. Here it was where a new advertising model called Advertising Network (AN) was born. The AN acts as an intermediary between advertisers and publishers, simplifying the process of launching campaigns on the Internet. The advertiser buys a number of impressions and the ANs displays that number of impressions on the pages of the publishers in exchange for a commission. Maximizing publishers' spaces while improving the performance of advertisers' campaigns are the main objectives of the ANs. These advantages led to the creation of numerous ANs. However, as time passed, and that, in spite of the increased demand for online campaigns, the number of ANs has been decreasing because large networks captured large market shares \cite{robinson2007internet}. The small ANs found themselves in a difficult situation because they lacked an effective system for online fraud detection and because they received a low number of visits, which made them unable to offer targeted campaigns to advertisers.

In addition, for advertisers, it is tedious to have to choose between the numerous ANs and evaluate their performance. For these reasons, currently, most advertisers have opted for large ANs because they offer more profitable campaigns \cite{goldfarb2011online}. In the same way, most publishers partner with large ANs because they obtain higher incomes and receive prompt payments. Having more advertisers and publishers make large ANs earn consistently higher revenues, which in turn allows them to offer more and better services to both advertisers and publishers. Large ANs can easily invest in promoting their platform and in recruiting new advertisers and publishers.

In this context, the smaller ANs decided to join, giving rise to a new advertising model called Real-Time Bidding. This new system networks can exchange visits to make a better match between users and the ads that are shown. Through RTB, small ANs can exchange adverts with each other in order to become more competitive. RTB consists of a large global market where publishers auction ad slots each time a user accesses to their web pages \cite{ren2019deep}.

Advertisers participate in the auction and the Generalized Second-Price (GSP) system is usually used to select the advertiser. In the system, the advertisement of the advertiser who makes the highest bet is selected and the price of the second highest bet is paid. Some studies show that this system is more favourable for publishers because they do not update the price constantly as in other systems \cite{cui2015global,yang2020learning}. However, maximizing economic performance in the short term does not guarantee that in the short-medium term it will be better.

In this paper, we propose an alternative payment model to GSP which takes into account not only the economic performance in the short term but also considers many other variables in order to guarantee that all involved parties (advertisers, publishers and specially ANs) will make reasonable profits.

The presented work is in line with that of Balseiro \cite{Balseiro2014}, in which he considers not only maximizing revenues but also the ad quality. The achievements of this paper consist of developing a RTB platform that evaluates when ranking an advert, all the indispensable requirements to make possible an adequate advertising ecosystem performance.

We consider our work to be of great interest due to the fact that it is the first article in RTB aimed at improving system performance by improving the ad selection system. The betting model presented takes into account many factors that are key to the proper development of the RTB advertising system.

The idea presented in this article can be adapted by the RTB networks, adding or removing some of the variables, making this model more beneficial for advertisers, improving their experience advertising and, therefore, attracting new advertisers that increase the volume of business of these platforms.

The rest of the paper is organized as follows. In Section 2, some related studies about RTB and bidding methods are presented. Section 3 explains the proposed method in general terms and illustrates the structure and each of the modules that compose the RTB platform, especially the ad selector module (ASM). Then, the main objectives for the proper functioning of the RTB platform, the rules to prevent online fraud and the penalties to ensure that the common objectives are met, are defined. Lastly, a methodology to optimize the weights of the ad selection function through a GA. In Section 4 our experiments are described and a brief analysis of the results obtained are drafted. In Section 5, the conclusions from our paper and some possible lines of research for future work are presented.

\section{Discussion of related works}

\subsection{Origin and structure of the Real-time bidding networks}

By working independently, most ANs waste those visits that do not match any of each individual advertiser's requirements. On the contrary, when the ANs share information about the users as well as when they exchange adverts, they increase the performance.

For example, if a user has left a product in a virtual store but has not yet paid for it, displaying virtual adverts will significantly increase the likelihood that the purchase will end. If such a user, visits a new page that belongs to the same advertising platform, they will be able to continue displaying ads in the virtual store\footnote{For a better understanding, we can imagine that all the pages belonging to the same advertising platform work as a single domain ( quote large article RTB).}.

That is why in Internet advertising in recent years a new advertising model has been created that works as a large global market called RTB where many advertising networks exchange information to increase their performance \cite {muthukrishnan2009ad}. On the other hand, RTB not only improves ANs' economic performance but also can improve fraud detection, since networks are able to share information about fraudulent techniques applied by both advertisers and publishers. Moreover, ANs can also share information about other faced threats such as click-bots\footnote{Click-bots are malicious programs that are typically installed on the user's computer and generate clicks automatically to hurt the advertising ecosystem \cite{miller2011s}.} and click-farms\footnote{The clicks-farms are groups of people who are trained to generate clicks on adverts in a given campaign in order to harm the owner \cite{mann2006click}.} \cite{blizard2012click}.

Although the first RTB platforms were created in 2005, the number of advertisers and publishers has skyrocketed \cite{yuan2013real}. Evidence of this growth is the company DSP Fikisu that aims to have a total of 32 billion impressions per day in 2017 \cite{zhang2017managing}.

RTB can be thought of as a global auction mechanism where advertisers compete with each other to display their adverts \cite{yuan2013real}. In this great market, advertisers make an offer for displaying their adverts on the websites of the publishers. If an advertiser wins the bid, its advert is displayed instantaneously. In the real-time auctions, the whole process of auction, acquisition and ad display, takes place in the time a user loads a website (less than 100 milliseconds) \cite{yuan2013real,adikari2015real}.

RTB platforms consist mainly of four modules: Demand-Side Platform (DSP), Supply-Side Platform (SSP), Ad Exchange (ADX) and Data Exchange (DX), where each module is responsible for a series of very limited tasks \cite{yuan2013real,adikari2015real}.

To understand in a general way the role played by each of the models we made the following description: The multiple ANs are grouped into a single module called ADX with their respective advertisers and publishers \cite{muthukrishnan2009ad}. These networks exchange user information through a module called DSP that improves the match between the ad and the user by significantly increasing the performance of the set of networks. The DSP is responsible for supplying advertisers with impressions that guarantee good performance for their campaigns. For this purpose, it uses algorithms that calculate the CTR of each impression. The main purpose of the DSP model is to offer a price in each auction that is carried out. On the other hand, the SSP tries to maximize the economic performance by managing the ad slots of the publishers in the most efficient way possible. This module distributes the information of each impression among as many networks as possible and selects the one that offers the higher price.

\subsection{Online campaigns optimization}

The increase in the volume of advertising in Internet and the amount of data that can be collected from users has made this area very interesting from the point of view of research. Many of the techniques developed in Data Science have been successfully applied to this type of advertising.

For example, some studies encourage ANs to apply the well-known machine learning techniques to online auctions \cite{perlich2014machine}. These models predict in a precise manner the acceptance of a user given an advert so that the probability of purchase increases considerably. In similar studies, adverts are ranked by the probability of being clicked, in such a way that the top-ranked adverts are likelier to be displayed \cite{cui2015global,edelman2005internet}.

In this line, new methodologies have arisen to improve auctions performance through the optimization of the user and website parameters so that the most interesting adverts are displayed. To this end, a simulation using GA and CTR estimation models takes place \cite{miralles2018novel}.

Supervised models and deep learning models are frequently applied for optimization in many situations related to online campaigns \cite{le2020overview}. For example, some publishers want to charge a fee regardless of whether users click or not on the advert, while some advertisers only want to pay if a click is generated. This problem can be solved using an intermediate role called ``arbitrageurs", and its success depends on how accurate the CTR estimations are \cite{cavallo2015display}.

Finally, an interesting approach related to the presented paper consists of focusing on optimizing advertisers' satisfaction rather than only considering economic performance. This is the premise that advertisers are more willing to do future investments in a particular advertising channel if they obtained a good performance. 

The Ad selection process can be seen as a combined optimization problem treated as a stochastic control problem. Policies for online advert allocation have been developed based on placement quality and advertisers' CPC bids \cite{Balseiro2014}. In this respect, the studies of Balseiro \cite{Balseiro2014} should be emphasized, since he makes a deep analysis of the balance that must exist between both economic performance by selecting the most profitable advert, and the quality of service offered to advertisers.

\subsection{Real-time bidding optimization approaches}

Generally, when we talk about optimizing RTB, we think about optimizing advertisers' campaigns through the price offered for each impression. This has been a recurring topic in the literature related to RTB. 

To not extend too much we will give two examples. First, Manxing Du, et al. \cite{du2017improving} increase the number of clicks of campaigns with low budget by applying a methodology based on Constrained Markov Decision Process, where the state is the estimated CTR, the action is the bid price and the reward is whether or not a click is generated. Second, Kuang-Chih, Lee et al. \cite{lee2013real} have another proposal to improve the RTB campaigns by selecting high-quality impressions. The idea they propose is that because conversions occur rarely and when they occur there is a delay in time, the most convenient is to spend the budget little by little to reach an audience as wide as possible. Along with this, each time an impression is displayed, its performance is evaluated to take into account in the future the impressions that improve the performance of the campaign.

In order to calculate the optimal price, the RTB methodologies are based on the predictions of models that estimate the CTR. So some research focuses on improving the accuracy of the models, which automatically improve the performance of betting strategies \cite{wang2017display}.

It is also possible to improve the performance of the RTB systems by optimizing the SSP module, that is, trying to maximize publishers' profits. In this sense, Shuai Yuan et al. \cite{yuan2014empirical} focus on fixing the reserve price or the floor price, which is the price below which the publisher is not willing to sell. Increasing this price means that, in some cases, the winners, instead of having to pay the price of the second highest bet, they have to pay the reserve price. In addition, in other cases, it will make some advertisers automatically raise their bets to get impressions. It is important not to raise the price too high since it could trigger the number of impressions that remain unsold.

Since Google introduced its Pay-per-click system in 2002, several studies have emerged in relation to online advertising payment methods. In the well-known Generalized Second-price (GSP) system, the highest bid wins the auction and the bidder pays a price equal to the second highest amount bidden \cite{cui2015global,edelman2005internet}. Even the GSP is not a verifiable auction system it continues to be one of the most implemented auction mechanism.

In this research, we propose to optimize the function to select an advertisement based not only on the economic aspect but we take into account a set of objectives such as the satisfaction of the publishers and the reduction of fraud (for more details see section 3) for the advertising ecosystem to work properly.

\section{Our novel Advertising Exchange System}\label{sec:Our-novel-Advertising}

The proposed AdX system implements an Advert Selection Function (ASF) that evaluates the necessary objectives for a proper system functioning. The objectives of our system are advertisers’ impression percentage, spam advertisers, campaigns profitability, advertising network balance, publishers’ click-fraud, and income maximization. These objectives are described in detail in subsection \ref{sub:DevelopmentAES}. It seems of most importance to us to develop a system aimed at the satisfaction of all the roles involved in online advertising rather than a system only focused on the selection of the most cost-effective advert.

In order to implement our AdX system, one variable will be used to represent each objective and one weight will be used to model each objective's importance in the advert selection formula, as expressed in equation \ref{eq:Ad_Rank}. The weights are optimized through a genetic algorithm (GA) according to the system's performance. The GA uses the system's performance, expressed in economic terms, as the fitness value. The fitness value is calculated by subtracting the total penalizations $(Pen_1,...,Pen_5)$ from the total income derived from the system. Our methodology is able to find the best values for the weights, given any configuration.

The best weights are those that maximize the income while minimizing the sum of all the penalties. Penalties are economic sanctions that are applied when a goal is not met. The less an objective is met the higher the associated penalty will be.

The value of these weights can be calculated offline and then the system configuration can be updated periodically. Our methodology is able to find the optimal weights using a GA from the definition of the objectives, the penalties, and the rules in order to prevent online fraud.

As it is shown in Figure \ref{fig:Advertising-Exchange-module}, in our proposed system, all ANs exchange adverts among themselves through the Advertising Exchange System (AES).

The most important AES processes are: selecting the best advert from among all the candidates, keeping the fraud detection system updated and managing collections and payments from advertisers and publishers \cite{korula2015optimizing,minastireanu2019light}.

\begin{center}

\begin{figure}

\centering{}\includegraphics[scale=0.28]{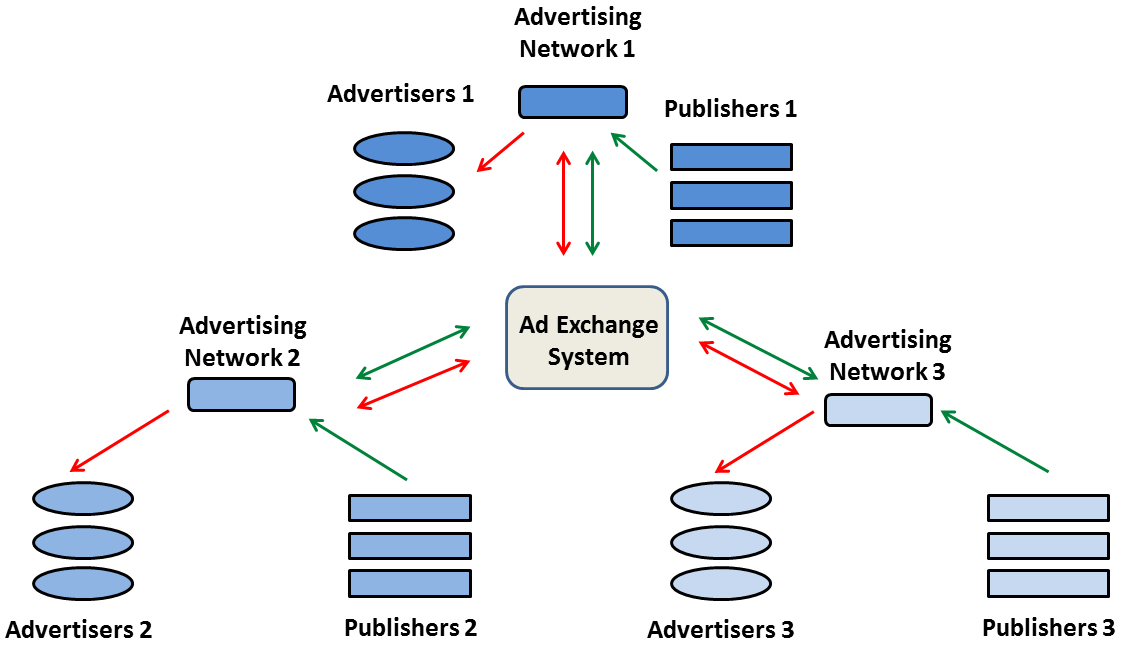}\protect\caption{\label{fig:Advertising-Exchange-module}Advertising Exchange System structure. The AdX consists basically of the AES and all the ANs that take part in the exchange of adverts.}

\end{figure}

\par\end{center}

\subsection{Advertising Exchange System}

In order to develop the AdX, we propose the AES shown in Figure 
\ref{fig:Advertising-Exchange-System}. The
designed AdX uses the CPC payment model\footnote{The CPC is the most widespread payment 
model and using several payment systems would greatly complicate the problem \cite{fain2006sponsored}.}. It is composed of four interconnected and interdependent modules: the CTR estimation module, the Fraud Detection module, the ASM and the database. Each module is designed for a different purpose and all of them are needed to make the advertising exchange possible.

The most important module is the Advert Selector. The other three
modules (CTR estimation, Fraud detection and Database module) provide the necessary information so that the Advert Selector can choose the advert with the highest performance.

\begin{center}
\begin{figure}
\centering{}\includegraphics[scale=0.34]{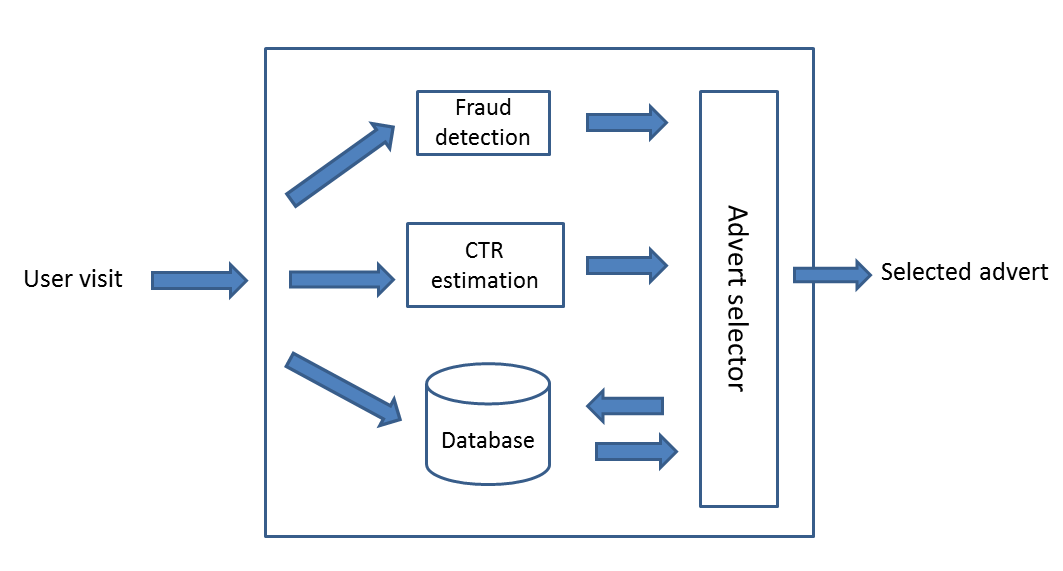}\protect\caption{\label{fig:Advertising-Exchange-System}Advertising Exchange System
structure. The AES is the cornerstone of our system since it performs all the
necessary functions for an appropriate advert exchange.}
\end{figure}
\par\end{center}

\subsubsection{Module 1: CTR estimation}

The CTR of an advert is calculated as the ratio between the number of clicks and the number of
impressions. But in the case of a single visit, the CTR can be computed
as the probability that a user generates a click on the advert
displayed on a website. This probability is expressed as a real number lying within the range {[}0,1{]}. Accurately estimating the CTR of
an advert is one of the biggest challenges of online advertising \cite{fang2014predicting}.

Bearing in mind that we implement the CPC payment method in this system, the ANs need to give priority to the most profitable adverts in order to maximize their income.

Machine Learning has been applied with great success in classification problems such as image and sound recognition \cite{zhong2011bilinear}, digital forensics \cite{sayakkara2020cutting} or CTR estimation \cite{miralles2017methodology}. In the case of CTR estimation, the dataset for machine learning methods contains data fields with users' features and websites' features such as advert size, advert position
or category of the page. The output of the model is ``1'' when the user generates a click and ``0'' when the user does not generate a click. We should clarify that, rather than predicting the class model, what is predicted is the probability that the output belongs to the class ``1'' in the {[}0,1{]} range.

\subsubsection{Module 2: Fraud Detection}

The Fraud Detection module informs about the probability of an advert being spam and the probability of a click being fraudulent.
The Fraud Detection module is designed to measure the probability
of an advert being spam and the probability of a click on the advert of a publisher's website being fraudulent.

The probability of fraud in both cases can be expressed as a real
number within the range {[}0,1{]}, $r\in R$, $r\in[0,1]$. As we
have mentioned previously, calculating the probability of fraud is
a highly complex process. Therefore, it becomes very difficult to
determine when a person is committing fraud from a single click or
from a single advert impression. To assess whether an advertiser or
a publisher is cheating, it is necessary to evaluate a large enough
set of clicks or advert impressions\footnote{This is the case of Google, which does not give a payment to publishers until they reach a sufficient number of clicks that allows determining
with great precision whether the publisher is fraudulent \cite{singh2006fraud}.}.

Moreover, the models that determine the probability of fraud would have to take into account the historical publishers'
clicks and advertisers' impressions. In the case of
spam adverts, some information regarding advertisers should also be
considered, such as the duration of the advertisers'
campaigns, the type of products he/she advertises, the adverts' CTR or the users' behavior when the advert is displayed \cite{zarras2014dark}.

In the case of click-fraud, some publisher's features should be examined \cite{kshetri2019online}. Furthermore, data about the users who visited the page need to be collected. Some important factors involved in detecting click-fraud are IP distribution, most
visiting hours, publisher's CTR, countries
and cities with more visits to the page, the type of users who obtain
access\footnote{The access to a website can be direct, through a link on another page or through a search engine \cite{plaza2011google}.} and users' behavior before and after generating a click \cite{zhang2008detecting}. The probability $P$ of an advert or a publisher's
click $Ad_{i}$ being fraudulent can be expressed as $P(Ad_{i}|fraud)=\alpha$,
and $P(Ad_{i}|not\:fraud)=1-P(Ad_{i}|fraud)$.

\subsubsection{Module 3: Database for algorithm execution}

The database records all the necessary information to carry out all the processes involved in online advertising. The database stores all the required information about advertisers, publishers, and ANs to allow the ASM to work optimally. The most important data stored in the database consists of information related to the advertisers' payments and the publishers' charges.

In addition, information about any fraud committed and information used by the ASF such as the advert CTR and the advert CPC fixed by each advertiser is also stored in the database. In the same way, whenever a user makes a visit to a page, an advert is displayed and the database is updated. The value of the probability that the click is fraudulent
and that the advertisement is spam is also updated.

\subsubsection{Module 4: Advertiser selection module}

Whenever a user accesses a publisher's website a selection
from among all adverts takes place. All adverts belonging to a different
category from that of the publisher's website that
is being accessed are discarded. Those adverts which are not discarded
are called candidates. Then, only one advert from among all the candidates
is selected, that is, the one that possesses the maximum $Ad\:Rank$
value. To select the best advert we apply the $\digamma(Advert)$
function, which assigns a real value in the range {[}0,1{]} to all
candidate adverts. The $Ad\:Rank$ is explained
in detail in subsection \ref{sub:Advert-Selector-Module}.

The $\digamma(Advert)$ function includes weights which are assigned
in proportion to the importance of each objective.
The $Ad\:Rank$ is calculated considering all the AdX objectives. As can be seen in Figure \ref{fig:Advertising-Exchange-System},
this module takes into account both the CTR and the likelihood of advertisers and publishers being fraudulent. It also consults and updates the database where information about advertisers'
campaigns, AN balance, publishers' account status and ANs' performance
is stored.

\subsection{\label{sub:DevelopmentAES}Development of the Advertisement Exchange System}

To develop the AdX we followed the following steps. First, we
defined the necessary objectives in order to ensure the proper functioning
of the publicity ecosystem. To ensure that objectives are met, we
defined one economic penalty for each objective, in such a way
that the more the objectives remain unmet, the greater the penalties sum
will be.

In addition, we created a set of rules in order to prevent the AdX from fraudulent activities. We established a metric expressed in economic terms in order to measure the AdX performance. Finally, we developed Algorithm \ref{alg:El-algoritmo-1} for the ASF and 
we defined a methodology to find the optimal configuration of weights using a GA.

\subsubsection{Definition of the objectives for the AdX}

Several objectives should be met in order to have a successful AdX \cite{cui2011bid} where
the optimization of some objectives may lead to the detriment of others. For example, the AdX should generate profits to the publishers as high as possible. But, at the same time, the AdX should not charge advertisers a price so high that their campaigns become unprofitable.

The objectives of the algorithm comprising all adverts $ad_{i}$ belonging to advertisers $adv_{i}\in Adv$, and all publishers $pub_{i}\in Pub$ of the $AN_{i}$, where $AN_{i}\in AdX$, are:

\begin{itemize}
\item \textbf{(O1) Advertisers' impression percentage:} All advertisers need to display a reasonable amount of adverts so that all of them are satisfied. If the algorithm focuses just on maximizing the income of the AdX, then some advertisers may be left with no impressions.
Thus, we should guarantee an equitable distribution of the advert impression number where advertisers paying a higher price have the advantage that their adverts are more frequently displayed.

\item \textbf{(O2) Spam advertisers:} Many advertisers display adverts on the Internet with malicious intent. These adverts are known as spam\footnote{Spam adverts sometimes redirect users to sites infected with a virus. They also make false products offers with a price below the market in
order to cheat users to obtain confidential data such as bank account
codes, email passwords or personal information. In other
cases, spam advertisers run a script to install malware or a Trojan
program on users' computers \cite{ellis2012web}.}. Spam advertisers are very detrimental to the online advertising ecosystem and so, we should calculate the probability that an advert is spam. We expect to reduce as much as possible the instances in which they are displayed. In case of implementing the system, we 
should also have a team in charge of verifying if an advertiser is 
trying to mislead users whenever the system alerts that an advertiser may be cheating.

\item \textbf{(O3) Campaigns profitability:} Some inexperienced advertisers
may pay for their campaigns a price above the prevailing market price.
It is not advisable to take advantage of this kind of advertisers
by charging them a higher price. Our AdX should make profitable campaigns
for all kinds of advertisers. Hence, we need to ensure that in our
AdX, the advert prices are similar to those in the market, that is $Prize_{ad}\backsimeq Price_{mkt}$.

\item \textbf{(O4) Advertising network balance:} Through collaboration, all ANs
should make it possible for other ANs to display adverts in other ANs. If
we want all ANs to participate in the AdX then the number of adverts received by each ANs should be similar to the number of adverts delivered, that is, $Adv_{rec}-Adv_{del}\backsimeq0$.

\item \textbf{(O5) Publishers' click-fraud:} Fraud committed by publishers is known as click-fraud and it can become very harmful to advertising campaigns. These fraudulent clicks are not generated by a genuine user interested in the product\footnote{They are performed with the intent of increasing the publishers' revenues or of harming the online platform. Many publishers may click on their own adverts or tell their friends to do so. There are also clicks made by click-bots which goal is to harm the advertising ecosystem \cite{blizard2012click,daswani2008online}.}. Due to click-fraud, advertisers end up paying for clicks that do not bring any benefit. This increases the likelihood that advertisers shift to another ANs offering more profitable campaigns. Thus, we should avoid displaying in the AdX spam adverts.

\item \textbf{(O6) Income maximization:} This is the most important goal,
but we place it in the last position because each of the previous
objectives has an associated penalty for it except this one. The Advert
Selector algorithm should look for the most profitable adverts in
order to distribute the highest amount of revenue possible among all
publishers. The income value represents the money collected from the advertisers.
Publishers should obtain reasonable economic returns so that they
are discouraged from moving to other platforms and encouraged
to recruit new advertisers.
\end{itemize}

\subsubsection{Economic penalties for the AdX}

To ensure that the objectives are met we define an economic penalty
$Pen_{i}$ and a coefficient $X_{i}$ associated with each penalty,
for each of the first five objectives $Obj_{i}$, where $i=1,...,5$.
In such a way that each penalty is applied whenever its corresponding
AdX objective is not met. The rationale behind these penalties is that those participants (ANs, advertisers, and publishers) who are not satisfied with the AdX usually
leave the platform, which translates into economic losses.
The $X_{i}$ coefficients allow us to increase or diminish the economic
penalization that is applied when a goal is not met.

The five penalties we have defined are:
\begin{itemize}
\item \textbf{(P1) Impression advert percentage:} We must apply a penalty
for each advertiser that fails to display a sufficient number of adverts.
P1 can be expressed as \emph{\textquotedblleft For
each advertiser whose average ratio of advert impressions lies below
25\%, we will subtract }$X_{1}$\emph{ times the average proceeds
of the advertisers in these ANs from the total Income''}.

\item \textbf{(P2) Spam advertisers:} We can define P2 as:\emph{ ``For
each click from a spam advertiser we will deduct }$X_{2}$\emph{ times
the money generated by these clicks from the total Incomes''}.

\item \textbf{(P3) Campaign profitability:} We want to avoid any abuse against inexperienced
advertisers who may be made to pay a price above the market price.
P3 can be expressed as \emph{ ``For each advertiser who pays a price 25\% above the market price for his/her campaign, we will deduct }$X_{3}$\emph{ times the money generated by that advertiser from the total Income''}.

\item \textbf{(P4) Advertising network balance:} When an AN is not satisfied,
it may stop working with the platform. Therefore,
P4 is expressed as \emph{ ``For each AN that receives 25\% fewer
adverts than the number of adverts it delivers, we will reduce $X_{4}$
times the incomes of that AN to the total Incomes''}.

\item \textbf{(P5) Publishers' click-fraud:} As mentioned
previously, click-fraud makes advertisers' campaigns
fail. To avoid this, we created the following penalty P5:\emph{
``For each fraudulent click from a publisher, we will deduct }$X_{5}$\emph{
times the value of this click from total Income''}.
\end{itemize}

\subsubsection{Online Fraud AdX Actions}

We should highlight that in our present study, fraud is
not just considered as an economic issue but also as an ethical issue.
Therefore, we must define a set of policies and rules oriented towards respecting their interests.

 \textbf{AdX Policies:}  Any publisher who wants to participate in the business must accept
several AdX policies aimed at reducing fraud to the greatest extent
possible, so that the advertising habitat may be protected. These
policies seek to expel publishers before they receive any payment
if the system's expert group determines that fraud
was intentionally committed. Additionally, we could consider imposing fines on advertisers who use the platform to deliver spam adverts and to all those publishers
who use black-hat techniques in order to increase their income.

\textbf{AdX Rules:} In addition to the AdX policies, we defined a
set of rules focusing on preventing fraud. These rules set clear-cut
criteria for expelling from the AdX those publishers, advertisers
or ANs who commit fraud. The difference between the rules and the
penalties is that infringement of rules leads to expulsion from the
AdX platform while penalties are used to undermine the performance when objectives have not been
met. In order to make the algorithm more efficient, we only check
the rules that lead to expulsion for each $N$ visits, where $N=1,000$.

The rules that we define are:
\begin{itemize}
\item \textbf{(R1) Fraudulent advertisers:} To dissuade advertisers from
trying to display spam adverts we defined the following rule: \textit{``If
an advertiser commits fraud on more than 20\% of his/her adverts and the
number of adverts is greater than 200 then he/she will be expelled''}
\item \textbf{(R2) Fraudulent publishers:} We expel those publishers whose malicious clicks amount to a certain percentage above a predetermined threshold $\mu$.
\textit{\emph{Hence, we defined the following rule: }}\textit{``If a publisher commits fraud on more
than 20\% of his/her clicks and the number of clicks is higher than a
specific threshold, in our case 30, then the publisher will be expelled''}.
\item \textbf{(R3) Fraudulent ANs:} To discourage ANs from allowing their
publishers and advertisers to commit fraud so as to win more money,
we defined the following rule: \textit{``}\emph{If 20\% or more of
the members of an AN are fraudulent advertisers or fraudulent publishers,
and the number of visits is greater than V, where $V=2,000$, then
the AN will be expelled from the platform}\textit{''}.
\end{itemize}

\subsubsection{\label{sub:Advert-Selector-Module}Advert Selector Module}

In order to optimize the performance of the algorithm tasked with
selecting an advert, we should define a function to evaluate
all the objectives defined above according to the pre-established economic metric. Since the system has six objectives, the ASF also has
six variables. Each variable is normalized and can
be expressed as a real number within the range {[}0,1{]}.

The weights assigned to each variable are represented by $\theta_{i}$,
in such a way that they satisfy the equation \ref{eq:weights}. These weights do
not have to be updated for each visit because this would lead to a
very high computational cost. The values of these weights can be recalculated
offline every few days. In addition, to ensure that the values of the weights are reliable, they must be calculated over a sufficiently large number of visits, since a small number of visits might not represent well the overall advert network behavior.

The weights' optimal value for a network may vary
depending on multiple factors such as the number of advertisers, the
number of publishers, the number of ANs, the average click-fraud and
the spam adverts within the AdX.

\begin{equation}
{\displaystyle \sum_{i=1}^{6}\theta_{i}}=1\label{eq:weights}
\end{equation}

To determine the best advert to be displayed on each user visit we
assign to each advert the $Ad\,Rank$ value. The $Ad\,Rank$
is recalculated for each candidate advert each time a user visits
a publisher's website applying the $\digamma(Advert)$ function as
expressed in equation \ref{eq:Ad_Rank}.

\begin{equation}
Ad\,Rank\gets\digamma(Advert)
\end{equation}

\begin{equation}
\begin{aligned}\digamma(Advert)=(\theta_{1}\times AN\,Satisfaction) +(\theta_{2}\times Advertiser\,Satisfaction)\\+(\theta_{3}\times Spam\,Adverts)+(\theta_{4}\times Campaign\,Cost)\\+(\theta_{5}\times Fraud\,Publisher) + (\theta_{6}\times Ad\,Value)
\end{aligned}
\label{eq:Ad_Rank}
\end{equation}

We now describe each of the variables representing the objectives
of the AdX system:
\begin{enumerate}

\item \textbf{AN Satisfaction:} It expresses the satisfaction of the members
of the network represented by the ratio between adverts received and
adverts delivered. We should give priority to the advertisers from the unbalanced networks. The closer
the value of this variable is to ``1", the more dissatisfied
are the members of the network. Hence, we should try to help those networks that are most dissatisfied. 
The values of the variables are normalized to the range {[}0,1{]} using equation \ref{eq:net-sat} to give priority
to unbalanced ANs.
\begin{equation}
AN\,Satisfaction=1-\frac{Received\,Visits}{(Received\,Visits+Delivered\,Visits)}\label{eq:net-sat}
\end{equation}

\item \textbf{Advertiser Satisfaction:} As expressed in equation \ref{eq:Satisfaction},
this variable measures the satisfaction of an advertiser according
to the number of impressions each advertiser obtains. The closer to ``1" the value 
of the variable is, the more discontent the advertiser will be. Therefore, we must give priority to those advertisers by displaying their adverts. 
\begin{equation}
Advertiser\,Satisfaction=\frac{Potential\,Visits}{(Potential\,Visits+Received\,Visits)}\,\times\,Ad\:Value\label{eq:Satisfaction}
\end{equation}

\item \textbf{Spam Adverts:} This variable represents the probability that an
advert is of spam type. The likelier an advert is to be spam,
the closer to zero the value of this variable will be. Therefore,
spam ads are less likely to be shown.

\item \textbf{Campaign Cost:} The price of a campaign must be similar to the
general market price. If an advertiser pays a price above the market price,
the value of this variable will get closer to zero, as expressed in
equation \ref{eq:cam-cost}.
\begin{equation}
Campaign\,Cost=\frac{Advertiser\:Price}{(Advertiser\:Price+Real\:Price)}\label{eq:cam-cost}
\end{equation}

\item \textbf{Fraud Publisher:} It represents the probability that a click is fraudulent. 
The likelier the publisher is to be fraudulent, the
closer to zero its value will be.

\item \textbf{Ad Value:} It represents the price the advertiser is willing to
pay and it is calculated by equation \ref{eq:ad-value}. The
closer to ``1'', the greater the price the advertiser
is willing to pay will be. To normalize the value of this variable
we divide the price the advertiser is willing to pay by the maximum
value of the category.
\begin{equation}
Ad\,Value=CTR\times\frac{CPC\:Advertiser}{Max(Category\:CPC\,Advertiser)}\label{eq:ad-value}
\end{equation}
\end{enumerate}

\subsubsection{Measuring the Advertising Exchange System performance}

In order to measure the AdX performance, we have established a metric
expressed in economic terms. As expressed in equation \ref{eq:AEM_Per},
the AdX performance is given by the difference between all the AdX
incomes and the sum of all the penalties. The algorithm tries to maximize
the AdX incomes, but at the same time it tries to achieve all the
objectives in order to minimize the AdX penalty value so that the
AdX performance value will be as high as possible.

\begin{equation}
AdX\,Performance=AdX\;Incomes\;\text{\textendash}\;AdX\:Penalties\label{eq:AEM_Per}
\end{equation}

The $AdX\,Incomes$ represents the money collected from all advertisers
from displaying their adverts, which is equal to the sum of the value
of all clicks, as expressed in equation \ref{eq:AEM-Incomes}.

\begin{equation}
AdX\:Incomes={\displaystyle \sum_{j=1}^{N}}Click\:Price\:(j)\label{eq:AEM-Incomes}
\end{equation}

$AdX\:Penalties$ is the sum total of all penalties, as expressed
in equation \ref{eq:Adx-penal}, and it represents the financial penalty
derived from not fulfilling the AdX objectives.

\begin{equation}
AdX\:Penalties=\sum_{i=1}^{5}Penalty\:(i)\label{eq:Adx-penal}
\end{equation}

\subsubsection{Mathematical system description}

Let us define a set of ANs as $ANs=<AN{}_{1},AN{}_{2},...,AN{}_{n}>$,
with $n$ number of ANs where each $AN{}_{n}$ has a list of advertisers
$Ad_{j}$ such that $\exists Ad_{j}\in AN{}_{n}$, a set of publishers
such that $\exists Pb{}_{k}\in AN{}_{n}$ and a set of visits such
that $\exists v{}_{l}\in AN{}_{n}$.

Each $Ad_{j}$ is defined by a set of adverts $Ad_{j}=<a_{1},...,a_{m}>$,
where $Ad_{j}\subseteq A$ and $(a_{i}\in Ad_{j}\wedge a_{i}\notin Ad_{m})$,
and A is the set comprising all the adverts. Finally, $V$ is the
set of visits $\forall v_{i}\in V;$ $\text{\ensuremath{\forall}}ANs$.

The selected advert $a_{i}$ is the advert belonging to the advert
set $A=<a_{1},...,a_{m}>$ and also $a_{i}\in Ad_{j}$ which leads
to the maximum income $I$, that is, select $A^{'}=\left\{ a_{i}|\,a_{i}\in A^{'}\wedge\:A^{'}\subseteq A\therefore a_{i}\in A\right\} $.
We must maximize the total Incomes $I_{k}$ and minimize the sum of
all penalties $P_{k}$ for all adverts $a_{i}$ from $AN_{k}$, that
is, $Max\left[{\displaystyle \sum_{k=1}^{N}}\left(I\,{}_{k}^{a_{i}}-P\,{}_{k}^{a_{i}}\right)\right]$
where $N$ is the number of ANs, $AN_{k}$ with $k=<1,...,N>$, for
an advert $a_{i}\in Ad_{j}$ and a $AN_{k}$ this system is subject
to:
\begin{itemize}
\item $Fraud\:(a_{i})>0:$ There is fraud on the part of the advertiser.
\item $Fraud\:(p_{i})>0:$ There is fraud on the part of the publisher,
where $p_{i}\in P$ and $P$ is the set of publishers.
\item $CTR\,{}_{k}^{a_{i}}=CTR\,{}_{k}^{a_{i}}\times\varphi_{j}$ and $\varphi_{j}$
represents the number of categories of $a_{i}$ with $\varphi_{j}\leq p$
where $p$ is the number of categories $C_{j}$ and $j=<1,...,p>\wedge\varphi\in\mathbb{R}$.
\item $CTR\,{}_{k}^{a_{i}}=\digamma{}^{a_{i}}(x_{1},x_{2},\text{\dots},x_{w})$
where $X=(x|\,x_{w})$ is an advertiser feature $a_{i}$.
\item $I\,{}_{k}^{a_{i}}=\left[\left(Click\times CTR\,{}_{k}^{a_{i}}\right)\times Price\,{}_{Click}^{a_{i}}\right]\times tc\,{}^{a_{i}}-(ep\,{}^{a_{i}}\times M\,{}^{a_{i}})$
where $tc$ is the total number of clicks on the advert, $ep$ is
the Income received by the publisher per click, $M$ is the number
of samples for the adverts and $P_{z}$ is the corresponding penalty,
and:
\end{itemize}
\begin{center}
$I\,{}_{k}^{a_{i}}={\displaystyle \sum_{i=1}^{6}\theta_{i}}$
\par\end{center}

\section{Calculating the optimal value of the weights using a Genetic Algorithm}
Each variable of the ASF represents one criterion and it is multiplied by a weight such that the sum total of all
the weights equals ``1'', as expressed in equation \ref{eq:weights}.
To obtain the optimal value for all weights, we applied optimization
techniques based on GAs.

Each time a visit occurs on a publisher's site within
the AdX, the ASM selects only one advert
from among the candidates.  Algorithm
\ref{alg:El-algoritmo-1} is in charge of taking into account all the objectives
and updating the variables used by the ASF. 

The optimal weight configuration is the combination that generates
the highest AdX performance according to the established metric. Algorithm \ref{alg:El-algoritmo-1} returns the AdX performance for a given weight configuration. We can think of Figure \ref{fig:M=0000F3dulo-intercambio} as a small module that returns the performance of the system (fitness of a GA function) according to the weights that are introduced as inputs. In order to find out the best weight configuration we apply a GA with the following components.

\subsection{Representation and initial population}

\begin{figure}
\centering{}\includegraphics[scale=0.42]{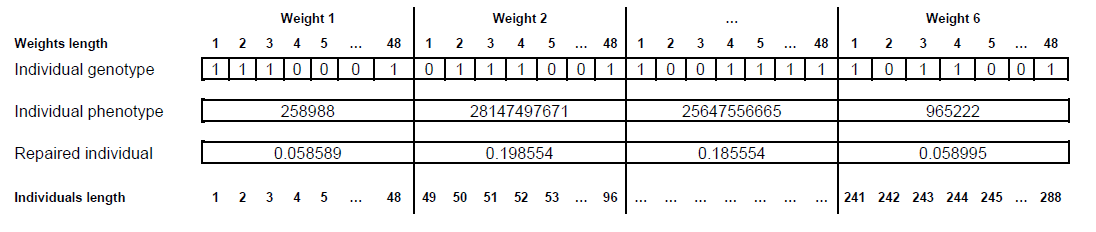}\protect\caption{\label{fig:poblacion}Weight codification using individuals of a GA.}
\end{figure}

As genotype, we use a binary fixed length representation. As it can be seen in Figure \ref{fig:poblacion}, we used a length of 48 bits to represent each weight. Therefore, each weight can be represented with a value between $0$ and $2^{48}-1$, which is a very high precision. Each individual $I$ of the population is formed by the six weights and it is represented by a string of $(6\times48\:bits = 288\:bits)$ binary digits. The initial population is obtained at random with a uniform distribution.

The size of the population is 100 in order to obtain diversity and an appropriate time of convergence \cite{golberg1989genetic}. The number of generations is 100 (Number of iterations in the stop criteria). Genetic algorithms have been used in optimization problems in various domains i.e. optimizing COVID-19 pandemic government actions \cite{covid2020miralles,covid2020optimization} .Therefore, the number of evaluations for the function goal is 10,000 ($100\:individuals \times 100\:generations = 100,000\: evaluations$). In some experiments, this number of assessments has been appropriate for the stabilization of the algorithm \cite{golberg1989genetic}.

\subsection{Handling constraints}

The genotype used to represent solutions does not satisfy the constraint that all weights add up to ``1''. However, an individual genotype $IG$ is a string of random binary digits can be converted in six integer numbers, which is called individual phenotype $IP$, where each integer is in the range ${[}0,2^{48}-1{]}$.

Once the individual $IG$ has been decoded into the individual $IP$, it can be easily transformed into a new array, called repaired individual $IR$, that satisfies the constraint (all numbers are in the range [0,1] and add up to one) applying Algorithm \ref{alg:Restriction}.

The repaired individual $IR$ should be calculated as a prior step to the evaluation of the individual. In this way, the constraint is always satisfied without the need to design specialized operators for solution initialization, crossover or mutation.

\begin{algorithm}
\begin{algorithmic}
\Require Individual IP 
\Ensure Repaired individual IR
\State $Sum  \gets 0$
\For {$i=1\:to\:6$}
\State $sum=sum+IP[i];$
\EndFor
\For {$i=1\:to\:6$}
\State $IR[i]=IP[i]/sum;$
\EndFor
\State return $IR$
\end{algorithmic}

\protect\caption{\label{alg:Restriction}Repair algorithm.}
\end{algorithm}

\subsection{Fitness function}
To calculate the fitness of each individual of the population $I$, the following steps are performed:
\begin{itemize}
\item Obtaining the repaired individual $IR$ (array of 6 real numbers in [0,1] that satisfies the constraint) of the individual $IP$.
\item Calculating the fitness value using equation \ref{eq:AdX-Fitness}.
\begin{equation}
Fitness(IR)={\displaystyle \sum_{j=1}^{N}}Click\:Price\:_{IR}\:(j) - \sum_{i=1}^{5}Penalty\:_{IR}\:(i)\label{eq:AdX-Fitness}
\end{equation}
\end{itemize}

\subsection{Genetic Algorithm parameter configuration}

We use ``Double-point'' for the crossover operator, that is, we select two points among which the genes of the individuals are interchanged. The parameter ``Elitism percentage'' is set to 5\%. The parent selection method used is the ``roulette wheel'' (proportional selection and stochastic sampling with replacement). The replacement used method is ``Generational replacement'' in which a completely new population is generated with new children from existing parents via crossover and mutation.

We applied similar parameters to the simple design GA proposed in Goldberg et al. \cite{golberg1989genetic}. The main reason is that our GA entails a high selective pressure (in comparison with other techniques of selection and generation replacement are a binary tournament or replacing steady-state) that takes a reasonable convergence time for our available computing capacity \cite{Back94selectivepressure}.

Since we used a binary simple representation and the constraint management does not require specialized operators, we consider to be appropriate the crossing and the uniform mutation operators proposed in Goldberg et al. \cite{golberg1989genetic}.

To find the best combination values, the mutation probability, and the crossover probability
are tested in the first configuration, which uses 10 ANs, with values from 0.1 to 1 with increments of 0.1. Therefore, we try 100 different combinations as expressed in Table \ref{tab:Crossing-mutation}. To calculate the best combination we chose the best average configuration after executing the algorithm 10 times. Once the best combination is selected, we run the algorithm 30 times and then we calculated the average of the fitness function. The time required for each execution to take place is of approximately 14 minutes and 25 seconds.

\begin{center}
\begin{algorithm}
\begin{algorithmic}[1] 
\Require ($\sum_{i=1}^{6}\theta_{i}=1:$ values), Data: Advertisers, publishers, ANs and users 
\Ensure $Fitness$
\ForAll{$vis_{i}\in Vis$} 	\Comment For all visits
\ForAll{$adv_{i}\in Adv$}     \Comment For all advertisers
\If{$(Category\:(Visit) = Category\:(Advert))$} \Comment Advert value calculation Function
\State $Ad\;Value \gets \!	 \begin{aligned}[t]             &  F((\theta_{1}\,\times\,AN\:Satis) +(\theta_{2}\,\times\,Adv.\:Satis) +(\theta_{3}\,\times\,Spam\:Adverts) \\ &
+(\theta_{4}\,\times\,Camp\:Cost) +(\theta_{5}\,\times\,Fraud\:Publisher)
+ (\theta_{6}\,\times\,Ad\:Value))          \end{aligned}$
\EndIf
\If{($Ad\;Value > Max$)}  \Comment Selects the best advertiser among all possibles
\State {$Max  \gets Ad\;Value$}  
\State {$Selected\;Ad \gets Ad_j$}  
\EndIf
\EndFor 
\If{($Num\:(Visits)\;mod\;1000 = 0$)} \Comment For each 1000 visits update parameters
\State $pen_{i}\in Pen,adv_{i}\in Adv,an_{i}\in ANs \gets$  UpdateParameters() \Comment Updates all roles parameters
\State $Apply\:Rule\:1(pub_{i}\in Pub, Adv_j$)   \Comment It checks if there are cheats publishers and ejects them
\State $Apply\:Rule\:2(adv_{i}\in Adv, Adv_j$)   \Comment It checks if there are cheats advertisers and ejects them
\State $Apply\:Rule\:3(AN_{i}\in ANs, Adv_j$) \Comment It checks if there are cheats ANs and ejects them
\EndIf
\EndFor  
\State Calculate the value of the variables: Incomes, $Pen_1,Pen_2,Pen_3,Pen_4,Pen_5$
\State $Fitness \gets Incomes - (Pen_1 + Pen_2 + Pen_3 + Pen_4 + Pen_5)$
\State Return $Fitness$
\end{algorithmic}
\protect\caption{\label{alg:El-algoritmo-1}Advertising exchange system algorithm.}
\end{algorithm}

\par\end{center}
\begin{center}

\begin{figure}
\centering{}\includegraphics[scale=0.43]{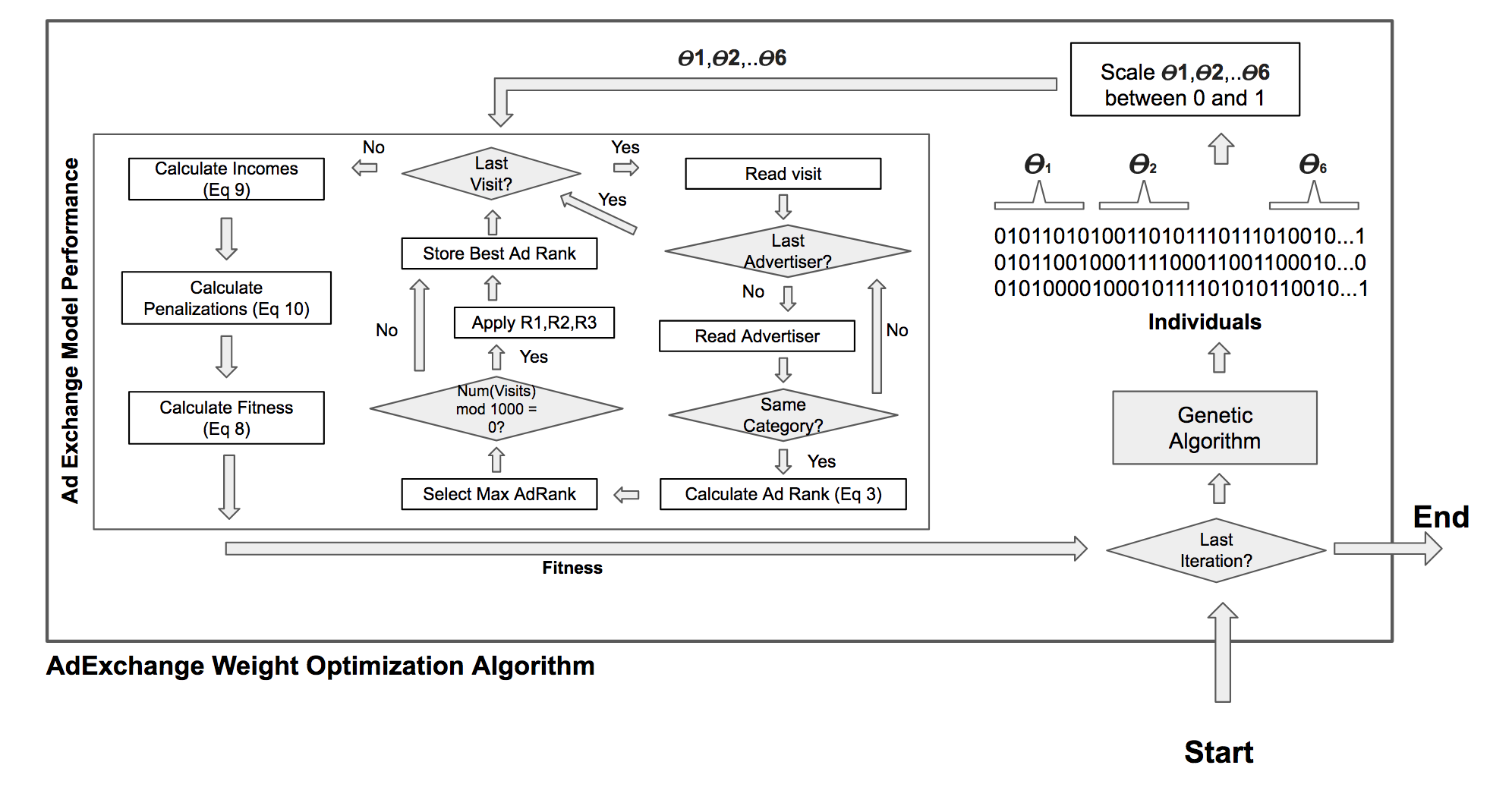}\protect\caption{\label{fig:M=0000F3dulo-intercambio}Advert exchange weight optimization algorithm using genetic algorithms.}
\end{figure}

\par\end{center}

\subsection{Justification for the chosen values of the coefficients, penalties and rules}

Click-fraud, spam adverts, and unsatisfied advertisers are factors that hurt the advertising ecosystem. However, determining the exact value of their negative impact on the AdX is a very complex task.
Even these values were calculated, we still cannot ensure that they will be optimal for a long time because the scenario could change quickly.

Therefore, finding the optimal configuration for all thresholds is out of the scope of our work and, for this reason, these values have been configured manually. However, we can briefly explain why we have configured the followings variables: 1) the coefficients $(X_1,...,X_5)$ associated with each penalty, 2) the thresholds above which penalties are applied and 3) the conditions of each rule to expel a role from the platform.

The thresholds of the penalties $Pen_1$, $Pen_3$ and $Pen_4$, representing the satisfaction degree, are configured to approximately 0.25\%. 

Penalties $Pen_2$ and $Pen_5$ refer to click-fraud and spam adverts, respectively. In penalties
$Pen_2$ and $Pen_5$, 1/2 times the revenue obtained by the fraudulent clicks and the spam adverts is subtracted to the total income.

With regards to the thresholds of the rules, we decided to expel from the AdX system all those ANs, publishers or advertisers committing more than 20\% of fraud. In order to decide if a party involved in the system has committed fraud, it is necessary to analyze a large enough set of
clicks or adverts.

In order to do this, we define the followings conditions. For publishers, the number of fraudulent clicks must be greater than 30. For advertisers, the number of adverts must be greater than 200. For ANs, the number of visits must be greater than 2,000. If instead of analyzing 150,000 visits, we analyze 10 million, the threshold values will have to be higher.

\section{Experiments and results}

To prove that our system is valuable, we compared in experiment I the performance of the GA system with the extended GSP method. After applying the GSP method, we applied the penalties defined in our system. Finally, the aim of experiment II is to demonstrate that our GA is capable of adjusting the values of its weights to the new system configuration.

\subsection{Preparation of the experiments}

Our system takes into account many parameters to select an advert such as spam adverts, CTR, fraudulent publishers, the bid price and so on. There are some data sets covering one of the considered aspects, but they are far from what we need. For this reason, to perform the experiments, both the visits and the configuration of each of the advertisers of all ANs have been simulated.

In this work, we launched an experiment that would help
us to understand the importance of each variable when the value of the
penalty remains constant. To find the optimum values of the weights,
we applied a GA. The GA is implemented in the environment Visual Studio C\# version 12.0.31101.00 Update 4, on a computer with the following features: Intel\textregistered Core i5-2400 CPU@3.10 GHz with 16 Gb RAM, with the operating system Windows 7 Pro, Service Pack 1, 64 bit.

We have used the Genetic Algorithm Framework (GAF) package for
C\#\footnote{The GAF is a .net/Mono assembly, freely available via NuGet, 
that allows implementing GA in the environment of programming
C\# using only a few lines of code \cite{Newcombe2015}.} to implement the 
GA. The GAF package was designed to
be the simplest way to implement a GA in C\#. It includes
good documentation and a variety of functions for crossover, mutation
and selection operators. In addition, it allows the customization
of the operator functions by the developer.

For achieving a deep evaluation of our proposed GA, we run the experiments I and II. We developed an environment of AdX with the following configurations. The percentage of an advert of being spam is randomly set within the range from 13\% to 16\%. The percentage of the publisher being fraudulent is randomly set with values in the range from 17\% to 20\%. The price the advertiser is willing to pay and the advert's real value are randomly set with values between 0.2 and 1.2 dollars. In the same way, the CTR value of an advert is randomly set in the range [0,1].

We used in experiments I the following number of ANs: 10, 20, 30, 40 and 50. Therefore, five different configurations are tested where each AN has 10 advertisers, 100 publishers and 150,000 user visits. Finally, each publisher’s page may belong to one of the 20 different categories and an advert can only be displayed in the pages with the same category.

In the first experiment, we compared the system performance both for the cases 
when ANs collaborate with each other and when they operate independently, 
by applying the famous GSP auction method \cite{chen2007optimal,edelman2007internet}. We conducted five configurations for the collaborative system and five for the independent system.

The GSP selects the advert with a higher price and the advertiser is charged with 
the value of the second priciest advert. Our system is focused on 
a collaborative AdX, so it would make no sense to apply the penalties when 
ANs operate independently. Therefore, we will not use the GA since 
there are no weights to be optimized in the ASF.

In this experiment, we have compared the profits obtained in the independent and
in the collaborative AdXs using the GSP methods. The average values of 30 executions are shown in Table \ref{tab:Exp-I}. The sum total of all income when ANs operate independently is 375,886.80\$ 
and 810,454.93\$ when they collaborate with each other. This is an increase of a 215.61\%.

When ANs work independently, the AdX displays only those adverts that belong 
to the AN which the user is visiting. However, when ANs collaborate with each other, 
adverts from any AN can be displayed.

If the AdX can choose an advertiser out of several networks instead of from only one, 
the results will be much better. As can be seen in Figure \ref{fig:GSP1},
the obtained profit when ANs collaborate is much higher than when they do 
not.

\begin{center}
\begin{table}[htbp]
\protect\caption{\label{tab:Exp-I} Results of the GA and the GSP systems.}
\centering { \noindent\adjustbox{max width=\textwidth}{%
\begin{tabular}{|l|c|c|c|c|c|}
\hline
\textbf{Nº of ANs} & \textbf{10} & \textbf{20} & \textbf{30} & \textbf{40} & \textbf{50} \\ \hline
Independent & 25,149.36 & 50,039.76 & 75,402.54 & 100,097.97 & 125,197.18 \\ \hline
Collaborative & 55,811.83 & 110,588.53 & 164,773.42 & 216,562.86 & 262,718.30 \\ \hline
\end{tabular}
}}
\label{}
\end{table}
\par\end{center}

\begin{center}
\begin{figure}
\centering{}\includegraphics[scale=0.60]
{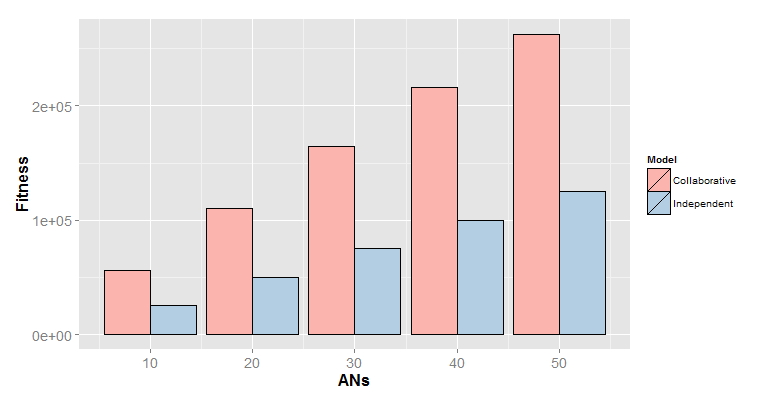}\protect\caption{\label{fig:GSP1}Experiment I: Obtained 
profits by the Independent and the Collaborative systems using five configurations.}
\end{figure}
\par\end{center}

\subsection{Experiment I}

In the second experiment, we configured the GA with the following settings. We set the coefficient value associated with each penalty as follows $x_{1}=x_{2}=x_{3}=x_{4}=x_{5}=0.5$. Assigning to all weights the same value allows us to see more clearly the relative importance of each objective.

These values are calculated by using the average value of ten different
experiments for each probability combination. As shown in
Table \ref{tab:Crossing-mutation}, the best probability combination
consists of a crossover probability of 0.7 and a mutation probability
of 0.2. Once we calculated the best combination, we executed the algorithm
30 times and we calculated the average. The results are shown in Table \ref{tab:Exp-II}.

\begin{center}
\begin{table}
\centering {\noindent\adjustbox{max width=\textwidth}{%
    \begin{tabular}{ @{} cl*{10}c @{}}         & & \multicolumn{10}{c} 
    {\textbf{Crossover prob.}} \\[2ex]         & & \textbf{0.1} & 
    \textbf{0.2} & \textbf{0.3} & \textbf{0.4}          & \textbf{0.5} & \textbf{0.6} & \textbf{0.7}  		& \textbf{0.8} & \textbf{0.9} & \textbf{1} \\  	
        \cmidrule[1pt]{2-12} & 
\textbf{0.1} & 9,920.4 & 9,971.5 & 9,711.7 & 9,997.3 & 9,783.6 & 9,763.8 & 10,018.1 & 9,893.0 & 9,750.1 & 9,898.5 \\ & \textbf{0.2} & 5,016.2 & 9,538.9 & 9,761.9 & 9,753.2 & 9,737.0 & 10,012.7 & \textbf{10,032.3} & 9,532.8 & 9,775.0 & 9,785.7 \\ & \textbf{0.3} & 9,509.8 & 9,757.2 & 9,810.6 & 9,630.8 & 9,804.0 & 9,808.1 & 9,493.8 & 9,803.0 & 9,693.1 & 9,606.2 \\ & \textbf{0.4} & 9,761.3 & 9,819.3 & 9,756.6 & 9,920.3 & 9,687.9 & 9,547.6 & 9,844.0 & 9,443.6 & 9,549.6 & 9,755.2 \\ & \textbf{0.5} & 9,828.0 & 9,561.0 & 9,625.4 & 9,454.0 & 9,633.1 & 9,710.0 & 9,743.5 & 9,873.1 & 9,365.4 & 9,629.7 \\ & \textbf{0.6} & 9,717.2 & 9,813.5 & 9,310.7 & 9,730.9 & 9,430.4 & 9,929.8 & 9,761.7 & 9,525.6 & 9,436.9 & 9,671.4 \\ & \textbf{0.7} & 9,507.1 & 9,604.4 & 9,569.9 & 9,691.2 & 9,565.6 & 9,490.1 & 9,532.3 & 9,878.3 & 9,297.7 & 9,255.0 \\
\rot{\rlap{~\textbf{Mutation prob.}}} & 
\textbf{0.8} & 9,932.8 & 9,776.1 & 9,212.0 & 9,417.7 & 9,513.3 & 9,724.2 & 9,738.0 & 9,312.8 & 9,410.1 & 9,825.9 \\ & \textbf{0.9} & 9,681.5 & 9,383.4 & 9,490.5 & 9,732.4 & 9,708.5 & 9,691.3 & 9,755.8 & 9,454.7 & 9,534.1 & 9,532.3 \\ & \textbf{1} & 9,609.7 & 9,479.9 & 9,788.1 & 9,716.4 & 9,630.7 & 9,609.4 & 9,977.5 & 9,383.0 & 9,893.3 & 9,947.2 \\
        \cmidrule[1pt]{2-12}
    \end{tabular} }}

\protect\caption{\label{tab:Crossing-mutation}Fitness value for Crossover and Mutation
probability for all possible crossover and mutation probability value combinations with 0.1 increments ranging from 0.1 to 1. These values are the average value of 10 executions.}
\end{table}

\par\end{center}

\begin{center}
\begin{table}[htbp]
\protect\caption{\label{tab:Exp-II} Average of the GA and the GSP systems for the five configurations.}
\centering { \noindent\adjustbox{max width=\textwidth}{%
\begin{tabular}{|l|c|c|c|c|c|}
\hline
\textbf{Nº of ANs} & \textbf{10} & \textbf{20} & \textbf{30} & \textbf{40} & \textbf{50} \\ \hline
GA & 10,146.59 & 19,188.95 & 30,861.78 & 41,587.55 & 50,167.97 \\ \hline
GSP & -26,727.29 & -41,331.81 & -63,645.60 & -100,379.85 & -124,853.94 \\ \hline
\end{tabular}

}}
\label{}
\end{table}
\par\end{center}

\begin{center}
\begin{figure}
\centering{}\includegraphics[scale=0.57]{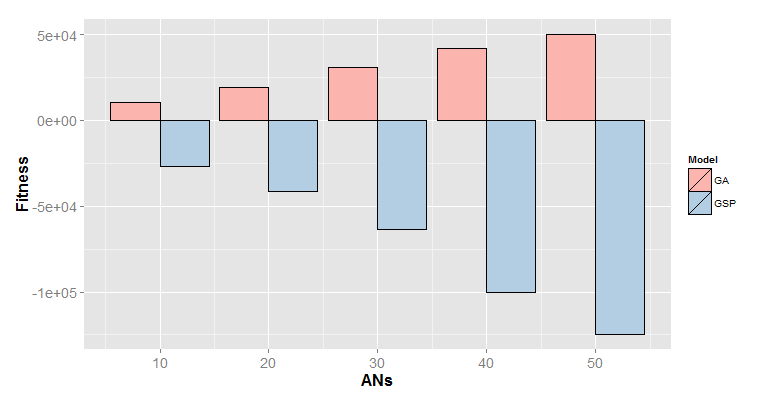}\protect\caption{\label{fig:GSP2}Experiment I: Comparison between GSP system and our GA system.}
\end{figure}
\par\end{center}

The optimal values of the weights in the first configuration, which uses 10 ANs, for the best fitness function are shown in Figure \ref{fig:Best-variable-configuration}. We have ordered
the variables in descending order according to their importance. 

\begin{center}
\begin{figure}
\centering{}\includegraphics[scale=0.45]{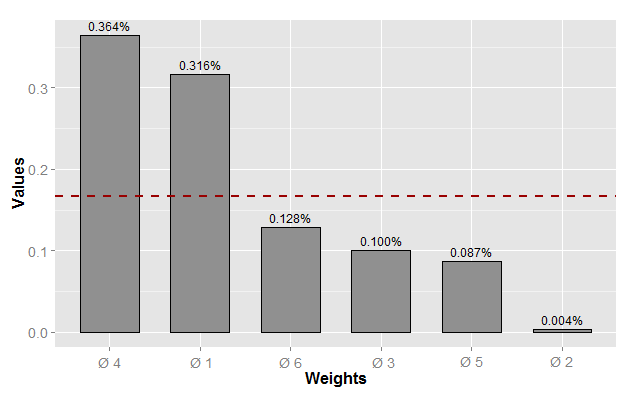}\protect\caption{\label{fig:Best-variable-configuration}Experiment I: Best weight configuration using 10 ANs. Each weight has a value lying within the range {[}0,1{]} and indicating the importance of each of the objectives. We have also represented the average of all the variables with a dashed red line.}
\end{figure}
\par\end{center}

As shown in Figure \ref{fig:GSP2} and in Table \ref{tab:Exp-II}, 
the performance of the GSP system is worse than the performance of our 
GA system. 

This is because the GSP system does not take into account any
objective defined for the AdX, but only the economic performance, and 
therefore the penalizations are very high. This makes us think that our 
system is interesting for those networks that want all their involved 
parties to be satisfied and want an ecosystem with little fraud.

As can be observed, weights $\theta_{4}$ and $\theta_{1}$ are the most important. We have to keep in mind that the metric used
in the fitness function is defined in economic terms. The weight $\theta_{4}$
is associated with $Campaign\,Cost$ and it indicates if an 
advertiser's campaign was priced above the market price. If those 
advertisers who are willing to pay more money for an advert were to leave 
the platform, the income would fall dramatically. 

On the other hand, $\theta_{1}$ regulates
the $Network\,Satisfaction$ variable which describes the network
satisfaction with respect to the number of visits received and delivered.
If a network leaves the AdX, all publishers and all advertisers who
belong to this network will be lost, and so the costs would be very
large.

$\theta_{6}$ represents the weight associated with the variable $Ad\,Value$,
which represents the advertisement value. It is logical that it should
have a high value because when more profitable adverts are selected,
the ANs' income increases. 

The weights $\theta_{3}$, $\theta_{5}$ and $\theta_{2}$ reflect
the values associated with fraud. $\theta_{3}$ is associated with
the $Spam\,Adverts$ variable, which indicates the probability that
an advertisement is of the spam type. $\theta_{5}$ is associated with
$Fraud\,Publisher$ which indicates if an advertisement is fraudulent.
Displaying spam adverts and receiving fraudulent clicks have a negative
impact on the AdX, this is why the value of these two
weights is similar. If we were to increase the value of these weights,
we would have to increase the coefficient $\theta_{2}$ associated
with this penalty value 2 or 3 times the amount of money obtained
through fraud, instead of just 0.5 times. 

Finally, weight $\theta_{2}$ is associated with $Advertiser\,Satisfaction$,
which indicates the satisfaction of an advertiser with respect to the 
number of adverts displayed. This weight usually has a value close to zero
and it leads us to think that it is almost of no importance, since
it is already automatically defined with weight $\theta_{4}$. This
means that, if the ANs are balanced, it is likely that the number of
adverts posted by the publishers will lie also within the objective set.

\subsection{Experiment II}

If we recall the results of Experiment I, we realize that  $\theta_{2}$ was the least weight on the optimization of the weights of each objective. In the following experiment, we are going to increase the weight associated with objective 3 to verify that the GA is able to adapt to these changes.

We made also another experiment with the same configuration as in this experiment, except for the value of the penalties' coefficients, in order to see how weight values are readjusted. 

To achieve this purpose, we create an experiment in which we
only change the coefficients of the penalties in the following way: $x_{1}=x_{3}=x_{4}=x_{5}=0.5$, while $x_{2}=3$, which represents the value associated with the
variable $\theta_{2}$. The rest of the parameters remains the same,
as in the configuration of experiment I. The average value of the 30
executions is 5,756.76. The results of the experiment
can be seen in Table \ref{tab:Exp-III}.

In this  system, we have used the  same configuration  as in the
previous system and we have also shown the calculated values.

Figure \ref{fig:Experiment-3} shows the results of the best weight
configuration with the highest fitness value. As it is shown, the most 
important value is $\theta_{2}$, which represents the advertiser's
satisfaction.

We can observe how the values of $\theta_{1}, \theta_{3}, \theta_{4}$
and $\theta_{5}$ continue to maintain the same order that they had
in Figure 1, in terms of their weight. This is obvious since all we
have done is to change the value of just one variable.

The conclusion is simple. We have shown that if we change the coefficients
of the penalties, then the values of the weights also change, so that
the advert selection formula is again optimized.

\begin{center}
\begin{figure}
\centering{}\includegraphics[scale=0.45]{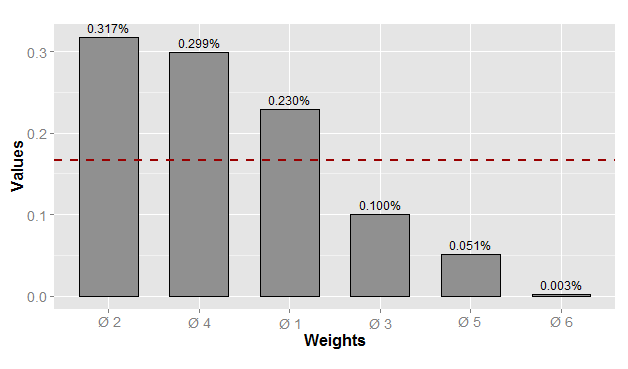}\protect\caption{\label{fig:Experiment-3}Experiment II: Best weight configuration changing the coefficients of the penalties.}
\end{figure}
\par\end{center}

\begin{center}
\begin{table}
\centering { \noindent\adjustbox{max width=\textwidth}{%
\begin{tabular}{|c|c|c|c|} \hline \textbf{Max value} & \textbf{Avg value} & \textbf{Min value} & \textbf{Std. dev.} \\ \hline 5,968.59 & 5,756.76 & 5,304.92 & 153.32 \\ \hline \end{tabular} 
}}
\protect\caption{\label{tab:Exp-III}Values of the genetic algorithm in experiment II.}
\end{table}
\par\end{center}

\section{Conclusions and future work}

Our work addresses a problem in the literature which,
although not much studied, is of no less importance.
To our knowledge, there is no other publication that focuses 
on creating a system for small networks to exchange adverts
among themselves in order to improve their performance.

We must bear in mind that the majority
of ANs do not reveal their algorithms and methods
since that would mean giving away part of their competitive advantage,
which may have taken them many years of research.

In this article, we have seen how to select an advert in an AdX system.
We have seen how the selection of an advert is not a trivial task 
but a complex task that must take into account multiple objectives,
often with conflicting interests, and each goal is associated
with a weight to be optimized.

One of the main achievements of this work is having provided a starting point from which an AdX system can be built and which takes into account the main threats and problems of online advertising.
In addition, a methodology was developed to find the appropriate weights for a function that considers all the necessary objectives that create a proper AdX ecoystem. 

Our goal was not to develop a methodology to improve CTR prediction
or fraud detection but to develop a methodology that helps in obtaining
the best advert selection function after assuming that the CTR and the Fraud detection modules were correctly developed. Obviously, the more reliable
and precise the modules that provide data are, the greater the system's
performance will be.

We have seen that the optimum weights for the advert selection module
vary depending on the goals, penalties, the number of advertisers and
campaigns, as well as the settings of everything that composes the
AdX. Therefore, there is no optimal configuration that can be extended
to all systems.

Studying the optimal value for each optimization would be an interesting
line for future works. These values could be found by constructing
complex simulated systems and testing them in a real scenario. 

As a future line of research, we might also attempt to include both
the CPM payment model and the CPA payment model to the AdX. Furthermore,
we may be able to develop new modules that enable ANs to cooperate
among themselves with the aim of improving fraud detection. For this
purpose, they could interchange information such as the CTR of the
page, the CTR of the adverts or the behavioral patterns of the users.

This could be done by collecting samples of behavior for later analysis
using models of machine learning. The more the samples and the greater
their quality, the more accurate can be the models that can be built. 

Further research could also involve developing a scalable system, i.e.,
instead of building a system of 10 networks with 10 advertisers and 100
publishers in each network, we could develop a system with 1,000 networks
comprising 10,000 advertisers and 100,000 publishers. However, this would require a
better hardware and more computers working in parallel.

Furthermore, to carry out this system, we could consider replicating the advert exchange system by using a distributed rather than a centralized architecture. These modules should be synchronized
with the ongoing exchange of information within the networks so that the variables are updated and so that they can optimize their response time for each user. In
order to do this, a communication protocol between the different Advert
Exchange Systems will be required. This protocol will transfer the
necessary information within the system in order to optimize the economic
profits of the system, avoid fraud and, finally, maintain the level of satisfaction of all the parties involved in the system.

\bibliographystyle{unsrt} 
\bibliography{GSP-RTB}

\end{document}